\documentclass[12pt,3p,preprint,sort&compress]{elsarticle}
\usepackage[colorlinks=true]{hyperref}

\usepackage{amsmath}
\usepackage{amssymb}
\usepackage{mathtools}
\usepackage[utf8]{inputenc}

\renewcommand{\d}{\partial}
\renewcommand{\l}{\left(}
\renewcommand{\r}{\right)}

\newcommand{\be}{\begin{equation}}
\newcommand{\ee}{\end{equation}}
\newcommand{\ba}{\begin{align}}
\newcommand{\ea}{\end{align}}
\newcommand{\bg}{\begin{gather}}
\newcommand{\eg}{\end{gather}}
\newcommand{\bseq}{\begin{subequations}}
\newcommand{\eseq}{\end{subequations}}

\begin{document}
\begin{flushright}
	INR-TH-2024-001
\end{flushright}

\title{NICA prospects in searches for light exotics from hidden sectors: the cases of hidden photons and axion-like particles} 
\author[inr,mpti]{Dmitry Gorbunov}
\ead{gorby@ms2.inr.ac.ru}
\author[inr,mpti]{Dmitry Kalashnikov}
\ead{kalashnikov.d@phystech.edu}
%\author[pnp]{Victor Riabov} 
%\ead{riabovvg@gmail.com}
\address[inr]{Institute for Nuclear Research of Russian Academy of Sciences, 117312 Moscow, Russia}
\address[mpti]{Moscow Institute of Physics and Technology, 141700 Dolgoprudny, Russia}
%\address[lpi]{P.N. Lebedev Physical Institute of the RAS, Moscow, Russia}
%\address[pnp]{
%Petersburg Nuclear Physics Institute named by B.P. Konstantinov of NRC Kurchatov Institute,   
%188300 Gatchina, Russia}
\begin{abstract}
    We present first estimates of NICA sensitivity to Standard Model extensions with light hypothetical particles singlet under the known gauge transformations. Our analysis reveals that NICA can explore new regions in the parameter spaces of models with a hidden vector and models with an axion-like particle of masses about 30-500\,MeV. Some of these regions seem unreachable by other ongoing and approved future projects. NICA has good prospects in discovery ($5\sigma$) of the new physics after 1 year of data taking.  
\end{abstract}
\date{}

\maketitle

%%%%%%%%%%%%%%%%%%%%%%%%%%%%%%%%%%%%%%%%%%%%%%%%%%%%%%%%%%%%%%%%%%%%%%%%%%%%%%%%
{\bf 1.} 
Nuclotron based Ion Collider fAcility (NICA) \footnote{\href{https://nica.jinr.ru/}{NICA website nica.jinr.ru}} is a new accelerator complex designed at Joint Institute for Nuclear Research in Dubna, Moscow region. The construction took about 10 years with active participation of thousands scientists and technicians from several tens of institutions all over the World. NICA plans operation with colliding beams of $p$, Au, Zn and the ion energy per nucleon will vary in the range 4-11\,GeV. 

Being a circular collider, NICA will operate with two collision points where two different instruments, Multi-Purpose Detector (MPD) \cite{Abgaryan_2022} and  Spin Physics Detector (SPD) \cite{Abramov_2021} are installed. While its main goal is investigating the hadronic matter --- nucleus structure, quark-gluon plasma, high baryonic density, etc --- it can be used for probing fundamental new physics which can show up in the heavy ion collisions. 

The kinematics implies that the most promising mass range for the new hypothetical particles to be produced by collisions at NICA is subGeV. The new physics, needed to explain  neutrino oscillations, dark matter, baryon asymmetry of the Universe, inflation, etc, can indeed be at such low energy scale, provided by only feeble interaction with known ingredients of  the Standard Model of particle physics (SM), for examples see correspondingly Refs.\,\cite{Asaka:2005an,Pospelov:2007mp,Akhmedov:1998qx,Bezrukov:2009yw}. Likewise, the large electric charge of the colliding ions favors models with the light particles coupled to photons as models to be explored to a larger extent. 

Hence, in this Letter we consider models with hypothetical light vectors (hidden photons) and models with axion-like particles (ALPs), both exhibiting direct interactions with photons. Physical motivation for these models, various concrete realisations, present constraints on the model parameters along with prospects of future experiments in testing their predictions can be found in e.g. recent reviews \cite{doi:10.1146/annurev-nucl-102419-055056, Feng_2023, berlin2023reviving}. Here we identify the dominant mechanisms of production of these hypothetical particles at NICA, their promising signatures, check them against the relevant background and finally estimate the NICA sensitivity to the model parameters. 

{\bf 2.} Both hidden photon and ALPs can travel macroscopic distances before decaying into visible particles. Given this fact we propose to search for the corresponding displaced vertex signature. The noticeable displacement reduces the otherwise too high background down to a negligible level. In recent MPD TDR an upgrade of Inner Tracking System (ITS) was proposed \cite{NICA_ITS}. This upgrade will allow one to determine the displaced vertex position with resolution up to $\sim 10$\,$\mu$m. The position of the ion collision (the actual point of each  collision) can be fixed within a region of about 100\,$\mu$m. To compare different possible outcomes we perform our calculations for three different minimal travel distances of new physics (NP) particles: $L_{min}=100$\,$\mu$m, $L_{min}=500$\,$\mu$m and $L_{min}=1000$\,$\mu$m. The probability of a particle to travel a distance larger than $L_{min}$ and nevertheless decay inside the detector is given by
\begin{equation}
    \label{Prob_1}
    P = \exp\l -\frac{L_{min}}{d} \r - \exp\l -\frac{L_{max}}{d} \r ,
\end{equation}
where $L_{max} \sim 1$\,m is the detector size, and $d=\tau \gamma \beta$ is the mean travel distance of NP particle. In our estimations we inferred $d_{max} \sim 1$\,mm that yields  $L_{max} \gg d$, $\exp\l -\frac{L_{max}}{d} \r \approx 0$, hence the probability\,\eqref{Prob_1} approaches  
\begin{equation}
    \label{Prob_2}
    P = \exp\l -\frac{L_{min}}{d} \r\,.
\end{equation}

In our work for the signal of $A'$ we consider only its electronic decay 
(i.e. $A'\to e^+e^-$), because in the present configuration the MPD is not equipped with muon detector, while the hadronic decay modes of $A'$ are undistinguished given the huge pion background. In the model with axion-like particle only the radiative decay (i.e. $a'\to \gamma\gamma$) is accounted as the potential signal source. However, there are other decays to SM particles. In each model we sum up all of them to the total decay width to SM particles denoted as $\Gamma_{SM}$. Moreover, in a particular model of the hypothetical particle there can be decay modes to particles from the hidden sector. Such unexplored decays can give sizeable contribution to the total decay width of the hypothetical particle $1/\tau\equiv\Gamma_{tot}$ and consequently $d$ and relevant branchings. To take into account these possible decays we additionally discuss the Half Hidden case (HH) with total width $\Gamma_{tot} = 2\Gamma_{SM}$ and the Half Hidden/5 case (HH/5) with $\Gamma_{tot} = 10\Gamma_{SM}$. The first case corresponds to models where the couplings to hidden sector are of the same order as the couplings to SM. The second case corresponds to models where the hidden sector decays dominate. The numbers of signal events then read 
\begin{align} \label{Signal}
    N_S & = N_{A'} \times P \times \frac{\Gamma(A' \rightarrow e^+e^-)}{\Gamma_{tot}}, \\
    N_S & = N_{a} \times P \times \frac{\Gamma(a \rightarrow \gamma \gamma)}{\Gamma_{tot}},
\end{align}
where $N_{A'}$ and $N_a$ are the numbers of produced hidden vectors $A'$ and ALPs $a$.

Light long-lived mesons like K-mesons are the main source of possible background, because these  mesons  also can travel macroscopic distances in the detector before their decay into visible particles. However, the maximum reach for NP mass at NICA appears to be less than kaon mass $m_K=498$\,MeV even in the background-free case. Another source of background is conversion of photons inside the detector volume. However, the current design of the ITS positions the initial elements of the detector at a distance of 20-30 mm from the beam axis \cite{NICA_ITS}. And dark photons are expected to decay well before reaching this distance, typically within a few millimeters from the beam axis. Given the high-vacuum conditions inside beam-pipe, which has approximately the same radii of 20-30 mm, we consider the contribution from photon conversion within inner region to be negligible. Still, pions from kaon semileptonic decays may be misidentified with electrons, which interferes with the suggested signature of $A'$. To avoid this type of background we follow the special procedure which singles out only pure (almost 100\%) electrons at the cost of 60\%-reduction of the expected electron statistics\,\cite{Abgaryan_2022}. Consequently, the background-free identification of the signal $e^+e^-$ pair will be allowed only for $0.4\times0.4=0.16$ of the total signal statistics. In what follows we adopt this procedure and to estimate the 95\% CL sensitivity of NICA to the model parameters we ask for the NICA operation time large enough to collect not 3 (as typically required by the Poisson statistics in the background free case) but 3/0.16=18 signal events.  

{\bf 3.} First, we investigate models with light vector hypothetical particle $A'$ coupled to the SM fields through the vector portal\,\cite{Holdom:1985ag} 
\begin{equation}
    \label{mixing}
    {\cal L}_{int}=-\frac{\epsilon}{2} \l \d_\mu A_\nu - \d_\nu A_\mu \r 
    \l \d_\mu A_\nu' - \d_\nu A_\mu' \r ,
\end{equation}
where $A_\mu$ stands for the SM photon field. This mixing with photon \eqref{mixing} yields $A'$ direct production in heavy ion collisions, where the big electric charge highly amplifies the production cross section with respect to $A'$ production in $e^+e^-$ collisions at similar energies, e.g. at future $c$-$\tau$ factories \cite{PhysRevD.107.015014, GORBUNOV2023138033,PhysRevD.100.115016,PhysRevD.99.015004}.

Likewise, one can take advantage of the multiple production inherent in hadronic scattering: the hidden photon can also emerge in photonic decays of mesons produced in heavy ion collisions. The most promising decays for the interesting previously unexplored range of $A'$ masses and mixing $\epsilon<10^{-3}$ are $\pi^0\to\gamma A'$, $\eta\to \gamma A'$ and $\omega\to \pi^0 A'$. The corresponding branchings are \cite{Ahdida_2021}
\begin{align}
\text{Br}\l P \to \gamma A'\r &= 2\epsilon^2\, \text{Br}\l P \rightarrow \gamma \gamma \r \times \l 1 - \frac{m^2_{A'}}{m^2_{P}} \r^3, \\ 
\text{Br} \l \omega \to \pi^0 A' \r &= \epsilon^2\, \text{Br}\l \omega \rightarrow \pi^0 \gamma \r \times \frac{\left[ \l m_\omega^2 - (m_{A'}+m_\pi)^2 \r \l m_\omega^2 - (m_{A'}-m_\pi)^2 \r \right]^{3/2}}{\l m_\omega^2 - m_\pi^2 \r^3}.
\end{align}
Since the parent mesons above are short-lived, all the hidden particles appear in the small vicinity of the ion collision point. The momentum distribution of the produced in this way $A'$ is obtained by boosting the isotropic distribution over 3-momenta in the meson decay frame to the laboratory frame. The boost parameters are determined by the transverse momentum $p_T$ and rapidity $y$ of the parent mesons, which we obtain by making use of the simulations of $10^5$\, Bi\,Bi collisions at central mass energy $\sqrt{s_{\text{NN}}} = 9.2$\,GeV per nucleon. The simulations are performed with generator PHSD for pseudoscalar mesons \cite{Cassing:2009vt} and with generator UrQMD for vector mesons \cite{Bass_1998}. 

In 1 year (50 weeks) of the MPD data taking at the differential luminosity of $L=10^{27}\,\text{ cm}^{-2}\,\text{s}^{-1}$ we get $N_\pi \sim 10^{13}$ neutral pions, $N_\eta \sim 10^{12}$ $\eta$-mesons and  $N_\omega \sim 10^{11}$ $\omega$-mesons \cite{NICA_wp}. Each channel contributes to the number of produced hidden photons $A'$ as ($X=\pi,\eta,\omega$) 
\[
N_{A'} = N_X \times \text{Br}\l X\to A'\r\,. 
\]

The produced hidden vector can potentially decay into the SM particles among which we recognize lepton and meson pairs with decay rates \cite{araki2023new} 
\begin{align}
    \Gamma\l A'\to l^+ l^- \r &= \frac{\epsilon^2 e^2}{12\pi} m_{A'} \l 1+\frac{2m_l^2}{m_{A'}^2} \r \sqrt{1-\frac{4m_l^2}{m_{A'}^2}},\\
    \Gamma \l A'\to hadrons \r&=\Gamma\l A'\to \mu^+ \mu^- \r \times R(m_{A'}) ,
\end{align}
where $R(\sqrt{s})=\sigma(e^+e^- \rightarrow hadrons)/\sigma(e^+e^- \rightarrow \mu^+\mu^-)$ is the energy dependent R-ratio \cite{pdg}.

\begin{figure}[!htb]  
\centerline{\includegraphics[width=0.5\linewidth]{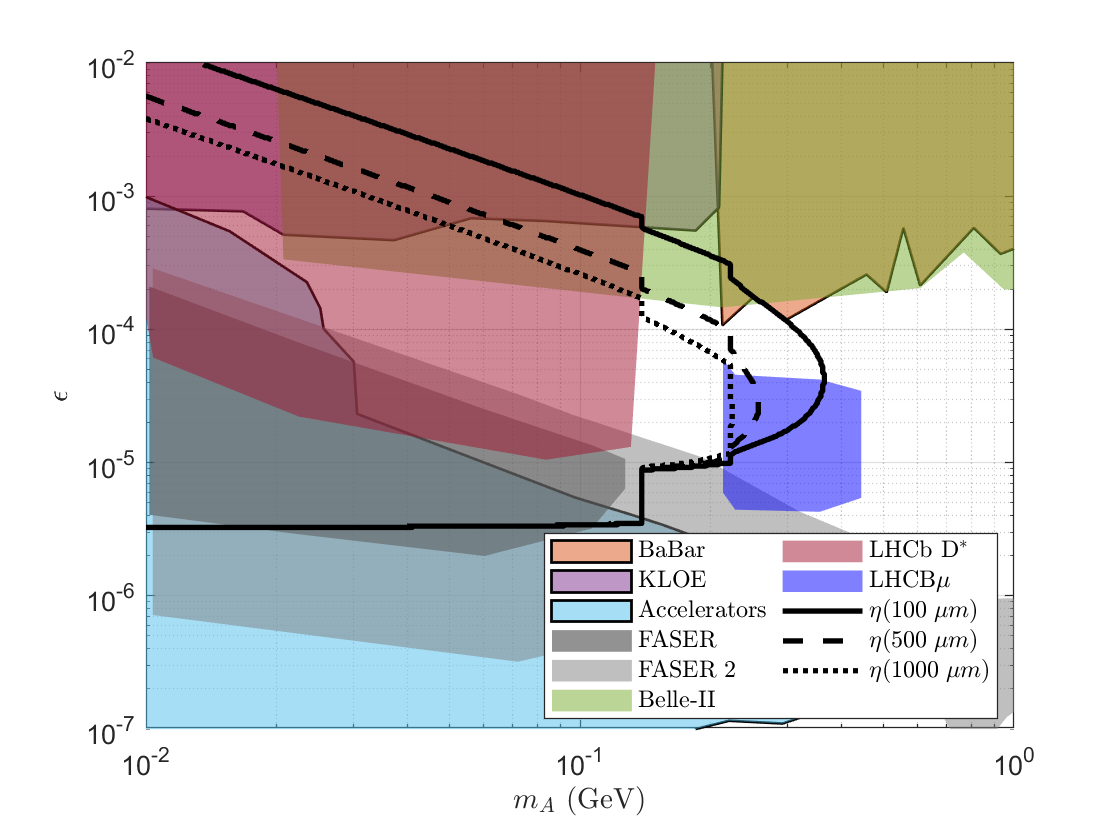} 
\includegraphics[width=0.5\linewidth]{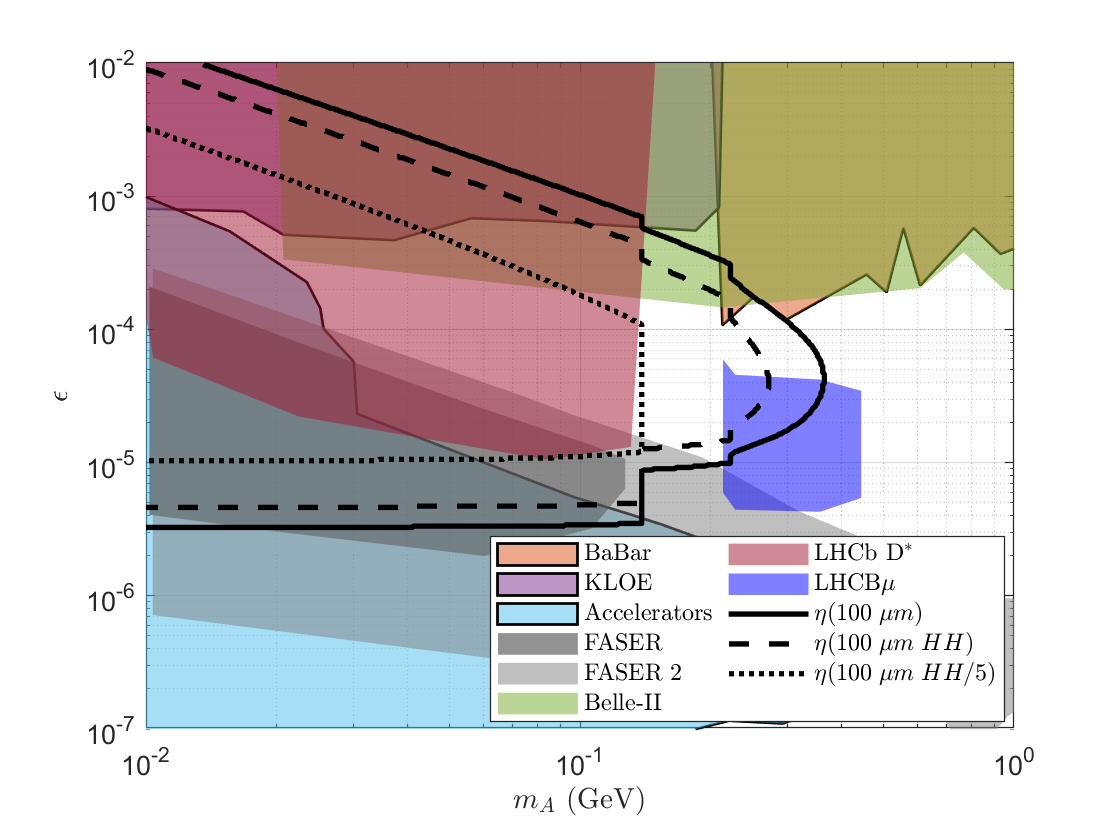} }
\caption{The regions to be probed at NICA after 1 year of operation at 95\% CL with $A'$ produced in $\eta$ meson decays. Left plot shows results for different minimal recognizable travel distance of $A'$, parameterized by $L_{min}$ in eq.\,\eqref{Prob_2}. Right plot shows results for different values of $\Gamma_{tot}$. The existing limits (colored and
outlined) are taken from BaBar at 90\% CL \cite{PhysRevLett.113.201801}, KLOE at 90\% CL \cite{ANASTASI2015633}, accelerator experiments (NA64 at 90\% CL \cite{PhysRevD.101.071101}, E141 at 95\% CL \cite{PhysRevD.86.095019}, NuCal at 95\% CL \cite{BLUMLEIN2014320}) and expected reaches of the ongoing experiments (colored) are given for FASER at 95\% CL \cite{PhysRevD.104.035012}, Belle-II at 90\% CL \cite{10.1093/ptep/ptz106}, LHCb D$^*$ at 95\% CL \cite{PhysRevD.92.115017}, LHCB$\mu$ at 95\% CL \cite{PhysRevLett.116.251803}.}
\label{fig:DP_from_eta}
\end{figure}

\begin{figure}[!htb]  
\centerline{\includegraphics[width=0.5\linewidth]{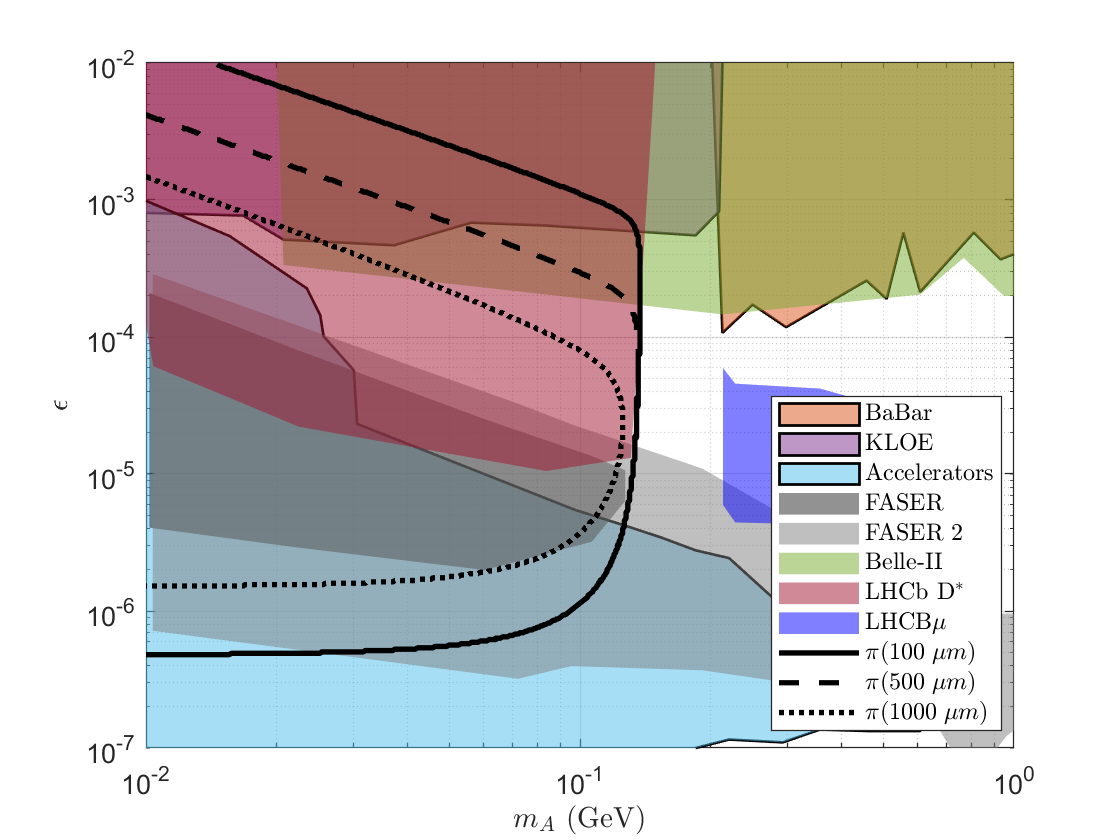} 
\includegraphics[width=0.5\linewidth]{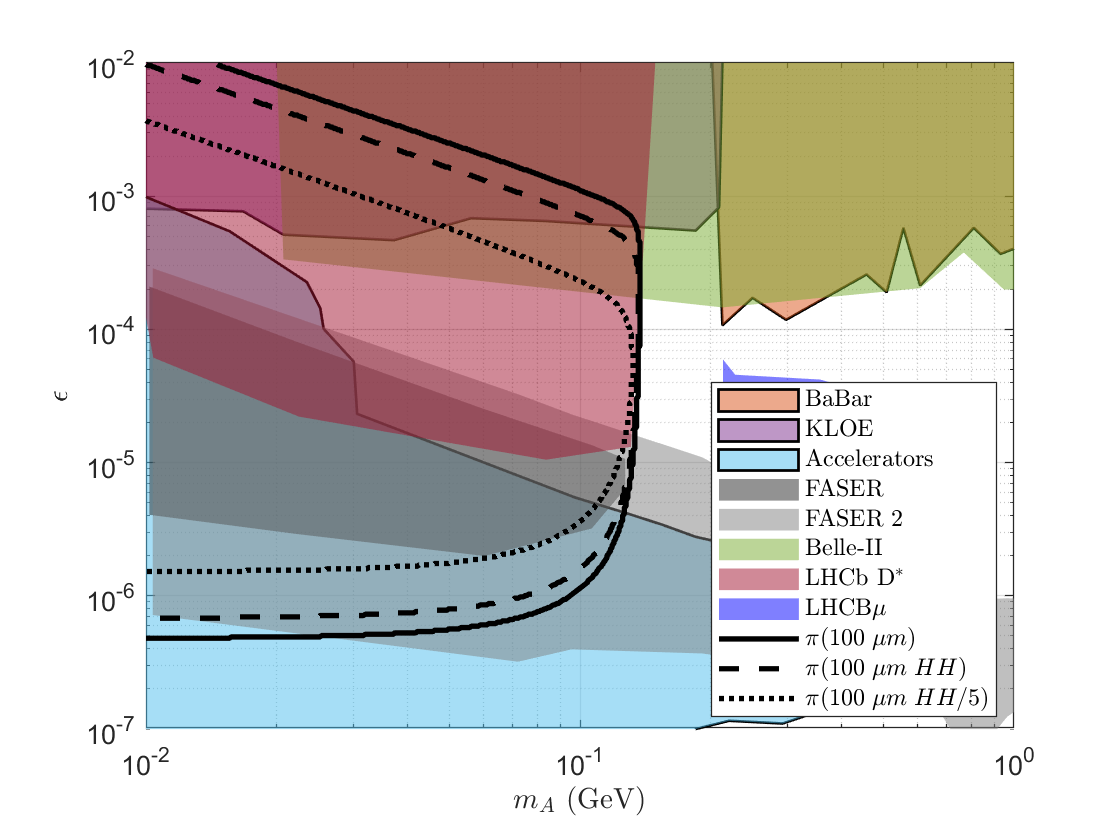} }
\caption{The regions to be probed at NICA after 1 year of operation at 95\% CL with $A'$ produced in $\pi^0$ meson decays. Left plot shows results for different minimal recognizable travel distance of $A'$, parameterized by $L_{min}$ in eq.\,\eqref{Prob_2}. Right plot shows results for different values of $\Gamma_{tot}$. The existing limits (colored and
outlined) are taken from BaBar at 90\% CL \cite{PhysRevLett.113.201801}, KLOE at 90\% CL \cite{ANASTASI2015633}, accelerator experiments (NA64 at 90\% CL \cite{PhysRevD.101.071101}, E141 at 95\% CL \cite{PhysRevD.86.095019}, NuCal at 95\% CL \cite{BLUMLEIN2014320}) and expected reaches of the ongoing experiments (colored) are given for FASER at 95\% CL \cite{PhysRevD.104.035012}, Belle-II at 90\% CL \cite{10.1093/ptep/ptz106}, LHCb D$^*$ at 95\% CL \cite{PhysRevD.92.115017}, LHCB$\mu$ at 95\% CL \cite{PhysRevLett.116.251803}.} 
\label{fig:DP_from_pi}
\end{figure}

\begin{figure}[!htb]  
\centerline{\includegraphics[width=0.5\linewidth]{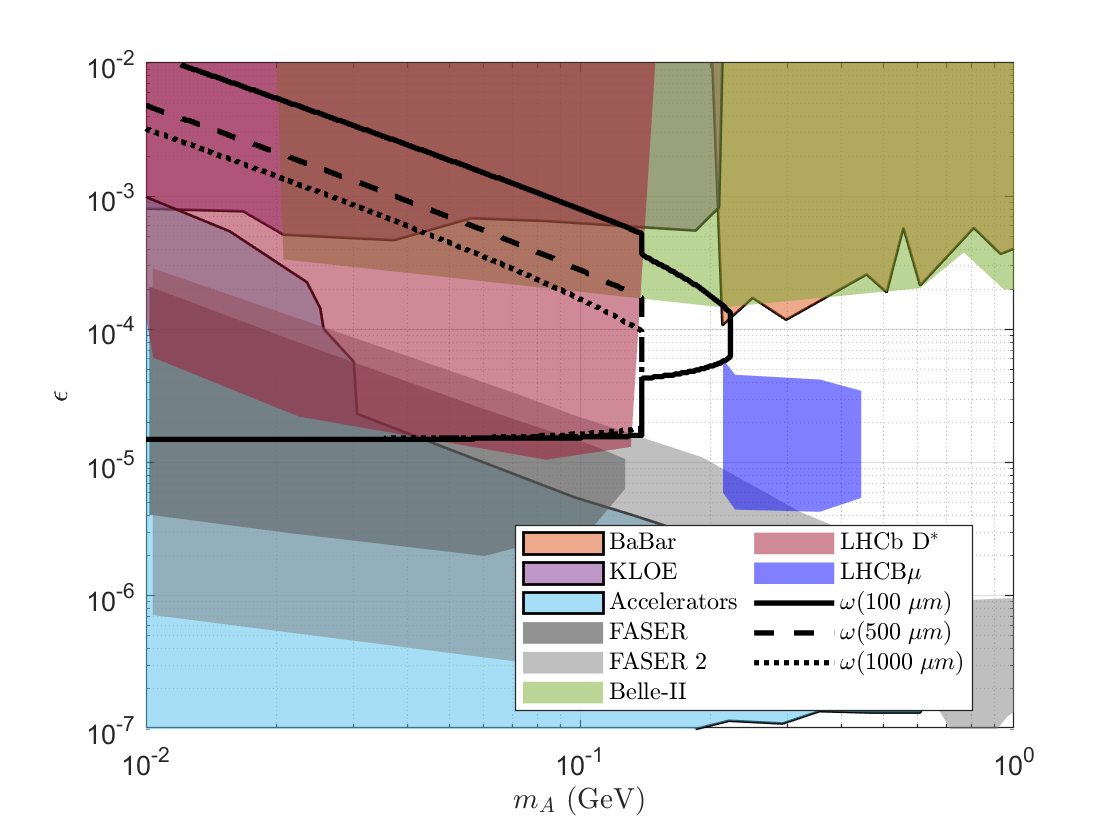} 
\includegraphics[width=0.5\linewidth]{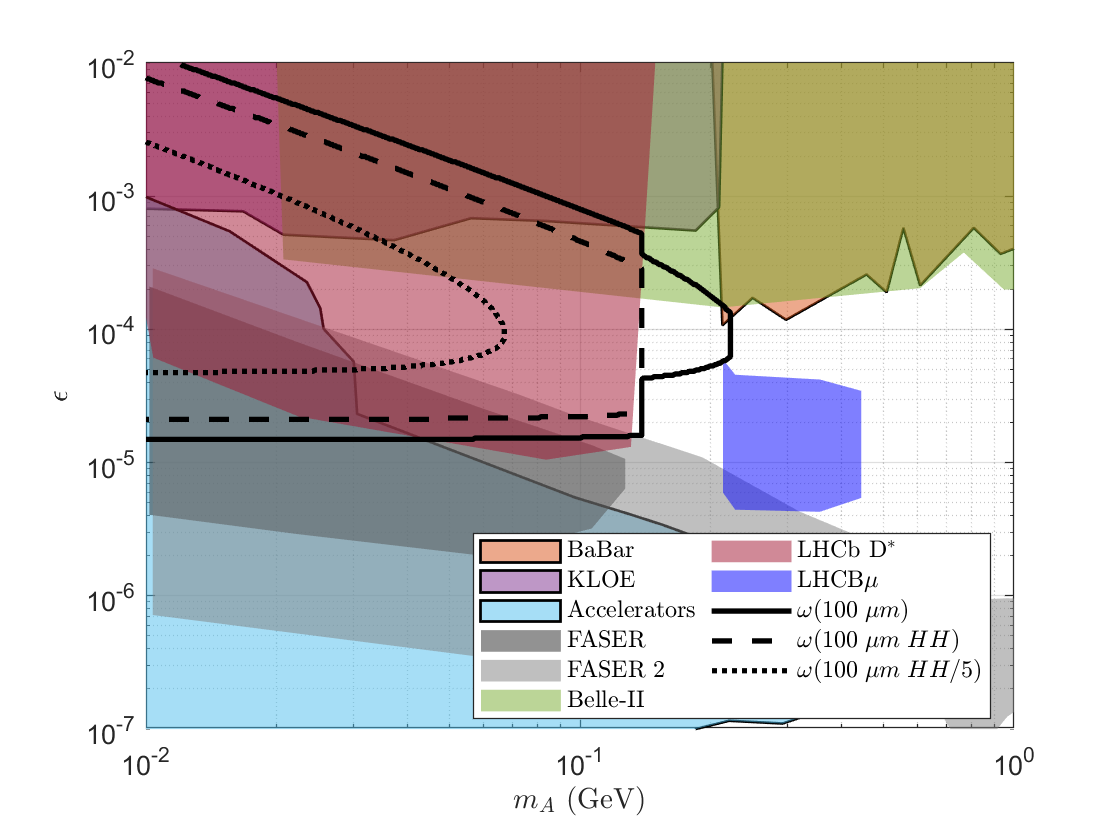} }
\caption{The regions to be probed at NICA after 1 year of operation at 95\% CL with $A'$ produced in $\omega$ meson decays. Left plot shows results for different minimal recognizable travel distance of $A'$, parameterized by $L_{min}$ in eq.\,\eqref{Prob_2}. Right plot shows results for different values of $\Gamma_{tot}$. The existing limits (colored and
outlined) are taken from BaBar at 90\% CL \cite{PhysRevLett.113.201801}, KLOE at 90\% CL \cite{ANASTASI2015633}, accelerator experiments (NA64 at 90\% CL \cite{PhysRevD.101.071101}, E141 at 95\% CL \cite{PhysRevD.86.095019}, NuCal at 95\% CL \cite{BLUMLEIN2014320}) and expected reaches of the ongoing experiments (colored) are given for FASER at 95\% CL \cite{PhysRevD.104.035012}, Belle-II at 90\% CL \cite{10.1093/ptep/ptz106}, LHCb D$^*$ at 95\% CL \cite{PhysRevD.92.115017}, LHCB$\mu$ at 95\% CL \cite{PhysRevLett.116.251803}.} 
\label{fig:DP_from_omega}
\end{figure}

Then we must account for the fact that the initial meson 3-momentum affects the decay length of $A'$ via non-unity boost factor. We integrate $(p_T,y)$ distributions obtained by simulations to get the normalized to unity $dn_X/dp_X$ distributions over meson 3-momentum $p_X$. Then we transform  Eq.(\ref{Signal}) accordingly and get for the number of signal events, these are electron-positron pairs coming from the vertex displaced from the ion collision region,
\begin{equation}
\label{signal-events}
    N_S = \int dp_X\, N_X\, \frac{dn_X}{dp_X} \times \text{Br}\l X\to A'\r \times \text{Br}\l A' \rightarrow e^+e^- \r \times \exp\l -\frac{L_{min}}{d} \r.
\end{equation}
Here $d\equiv 1/\Gamma_{tot} \times p_{A'}/m_{A'}$ and $p_{A'}$ is the $A'$ 3-momentum in the laboratory frame. Since MPD can only detect photons with energies exceeding $50$ MeV, we accordingly constrain the momenta in the integral\,\eqref{signal-events}. Within this approach we arrive at the results presented in Figs.\ref{fig:DP_from_eta}, \ref{fig:DP_from_pi}, \ref{fig:DP_from_omega}, 
where black lines correspond to 18 signal events (lepton pairs) consistent with limits at 95\% CL in the background free case according to our procedure explained above. 

On these plots we illustrate the NICA sensitivity to model parameters $\epsilon$, $m_A$ considering separately the three sources of $A'$ production. We also investigate the impact of the total decay width $\Gamma_{tot}$ of $A'$ on the NICA sensitivity. To this end for the plots on left panels we assume $\Gamma_{tot}$ to be a free parameter, but constrained in such a way, that $\Gamma_{tot}\gtrsim \Gamma_{SM}$ and corresponding particle decay length defined below Eq.\,\eqref{signal-events} is large enough to avoid the suppression \eqref{Prob_2} due to minimal recognizable decay length $L_{min}=100,\,500,\,1000$\,$\mu$m. The latter suppression guarantees, that the decay vertex is displaced from the ion collision point at the distance sufficient to be recognizable, and no any background is expected at such distances.  For the plots on right panels we assume particular relations between the total width $\Gamma_{tot}$ and the width into SM modes $\Gamma_{SM}$ and constrain the decay length $d$ from below by choosing the minimal displacement as $L_{min}=100\,\mu$m.   

One concludes, that new regions of the model parameter space can certainly be explored at NICA after 1 year of operation. The production of $A'$ is dominated by $\pi^0$ and $\eta^0$ decays. 

To estimate the contribution of the direct production of $A'$ in heavy ion collisions we consider its production in ultraperipheral collisions via $A'$-strahlung, assuming colliding ions remain intact. To calculate the cross section we make use of the CalcHEP package \cite{Belyaev_2013} and multiply each   electromagnetic vertex with fermions by the monopole form factor \cite{Vysotsky:2019qou}. The obtained results show that cross section of such processes is too small to produce interesting amount of $A'$.

\begin{figure}[!htb] 
    \centering
    \includegraphics[width=0.3\textwidth]{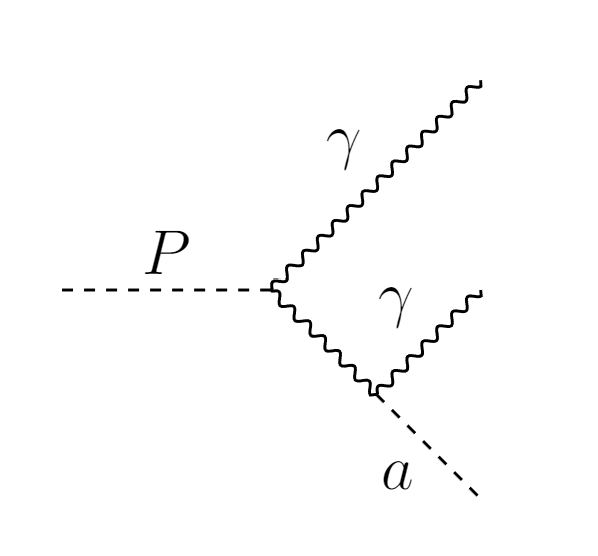}
    \caption{The Feynman diagram for the production of axion in pseudoscalar radiative decays.}
    \label{fig:Diagram_axion}
\end{figure}

{\bf 4.} The second model we investigate is Axion Like Particle (ALP) with coupling to photons 
\begin{equation} \label{axion-photon}
    \mathcal{L} = \frac{1}{8} g_{a \gamma \gamma} \; a \, F_{\mu \nu} F_{\lambda \rho}\epsilon^{\mu\nu\lambda\rho}.  
\end{equation}
Along with effective coupling of neutral pseudoscalar mesons $P=\pi^0,\,\eta,\,\eta'$ to photons \cite{pich1998effective},
\begin{equation}
    \mathcal{L} = \frac{\alpha}{4\pi f} c_P \; P \, F_{\mu \nu} F_{\lambda \rho}\epsilon^{\mu\nu\lambda\rho},     
\end{equation}
where  $f=92.4$\,MeV, $c_\pi=1, \; c_{\eta}=1.10, \; c_{\eta'}=1.34 $, eq.\,\eqref{axion-photon} yields decays of mesons to axions $P \rightarrow \gamma \gamma a$ through the diagram depicted in Fig.\ref{fig:Diagram_axion}.

The width of ALP decay to photons is
\begin{equation}
    \Gamma(a \rightarrow \gamma\gamma) = \frac{g_{a\gamma\gamma}^2m_a^3}{64\pi}.
\end{equation}

In our analysis, calculating ALP production and ALP decays we account for only ALP coupling to photons. Generically, there also could be ALP interactions with leptons and quarks, as we have in QCD axion models. Couplings to fermions open new decays to pairs of electrons, muons and mesons (if kinematically allowed). Typically, ALP decay rates to pair of leptons are suppressed by $m_l^2/m_a^2$ \cite{Liu_2023}. Nevertheless one may say we somehow account for them introducing HH and HH/5 cases with an additional to photon contribution to the ALP total width. Coupling to fermions could also provide with another sources of ALPs like weak decays of $K$-mesons \cite{Cortina_Gil_2021, guerrera2021revisiting}. In our case, where the photon coupling dominates, we neglect all other possible additional sources of ALP production, and consider only two-photon displaced vertex as the signature of ALP decay inside the NICA detector. It implies, that together with HH and HH/5 cases we obtain conservative estimates of the NICA reach.

\begin{figure}[!htb]  
\centerline{\includegraphics[width=0.5\linewidth]{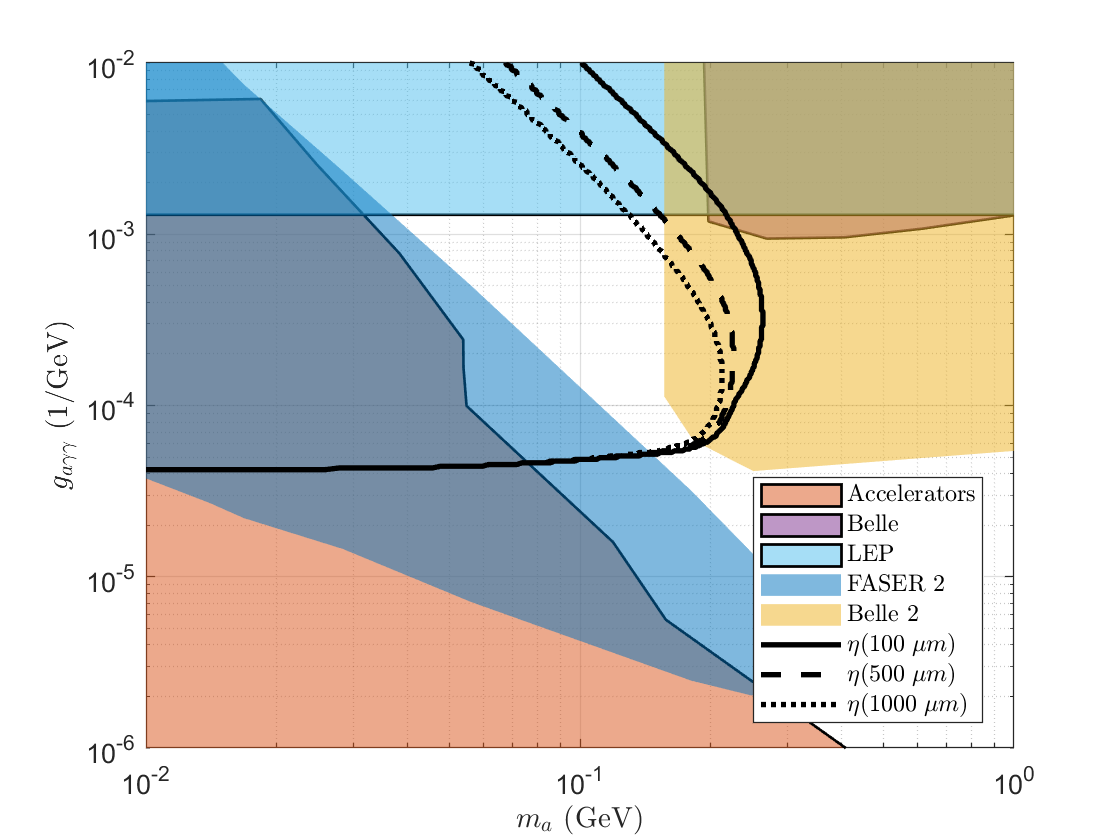} 
\includegraphics[width=0.5\linewidth]{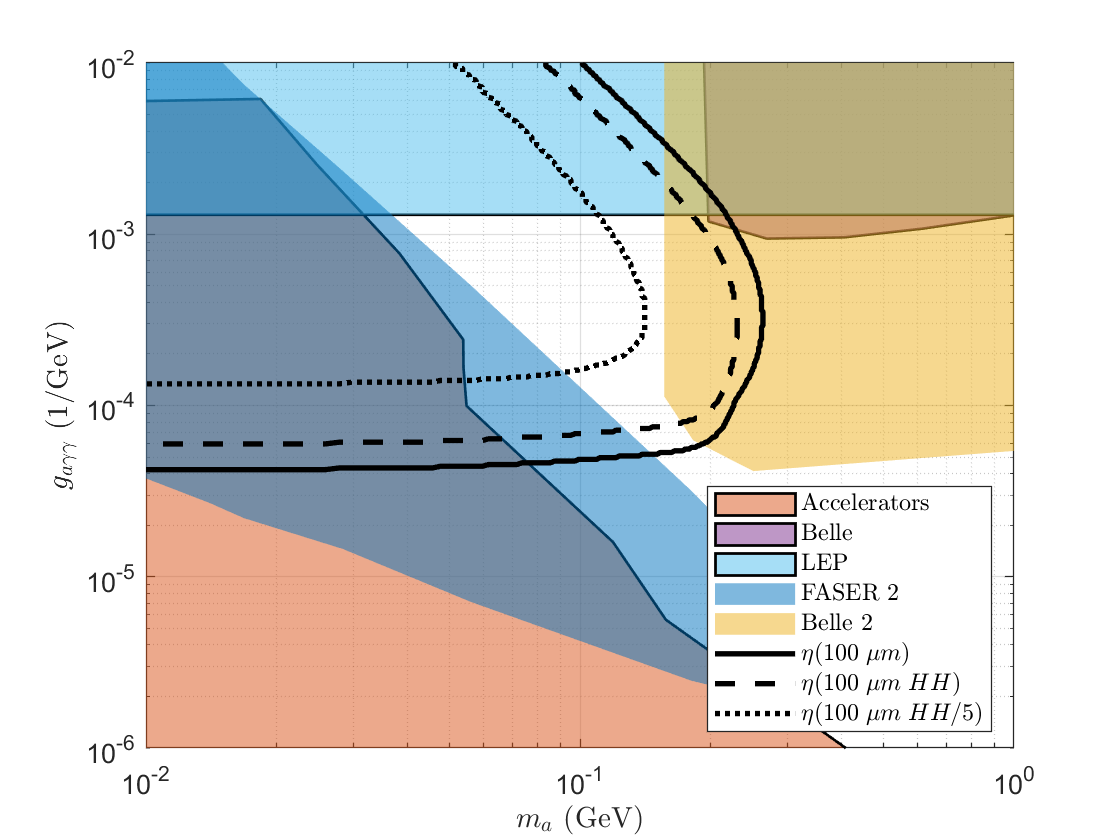} }
\caption{The regions to be probed at NICA after 1 year of operation at 95\% CL with ALPs produced in $\eta$ meson decays. Left plot shows results for different minimal recognizable travel distance of ALP, parameterized by $L_{min}$ in eq.\,\eqref{Prob_2}. Right plot shows results for different values of $\Gamma_{tot}$. The existing limits (colored and
outlined) are taken from Belle at 95\% CL \cite{PhysRevLett.125.161806}, LEP at 95\% CL \cite{LEP}, accelerator experiments (NA64 at 90\% CL \cite{PhysRevLett.125.081801}, E137 at 95\% CL \cite{PhysRevD.38.3375}, NuCal at 90\% CL \cite{NuCal_alp}) and expected reaches of the ongoing experiments (colored) are given for FASER at 95\% CL \cite{PhysRevD.98.055021}, Belle-II at 90\% CL \cite{Dolan2017}.} 
\label{fig:Axion_from_eta}
\end{figure}

\begin{figure}[!htb]  
\centerline{\includegraphics[width=0.5\linewidth]{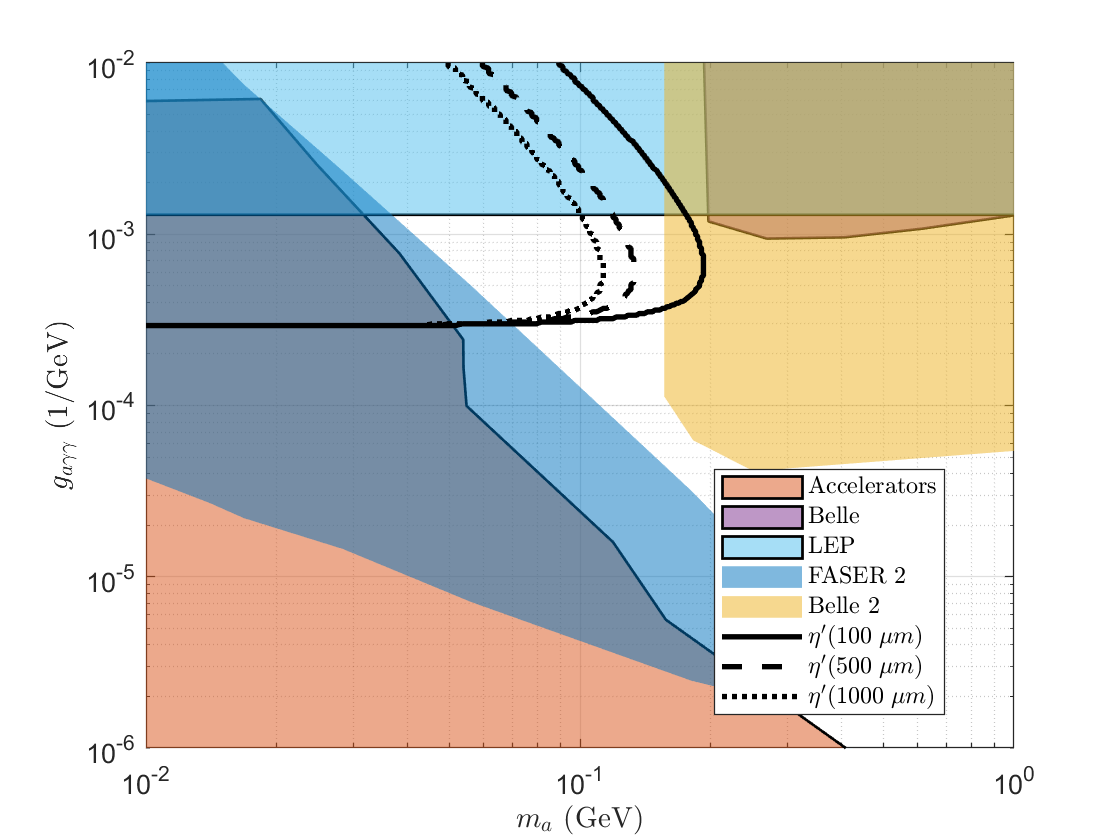} 
\includegraphics[width=0.5\linewidth]{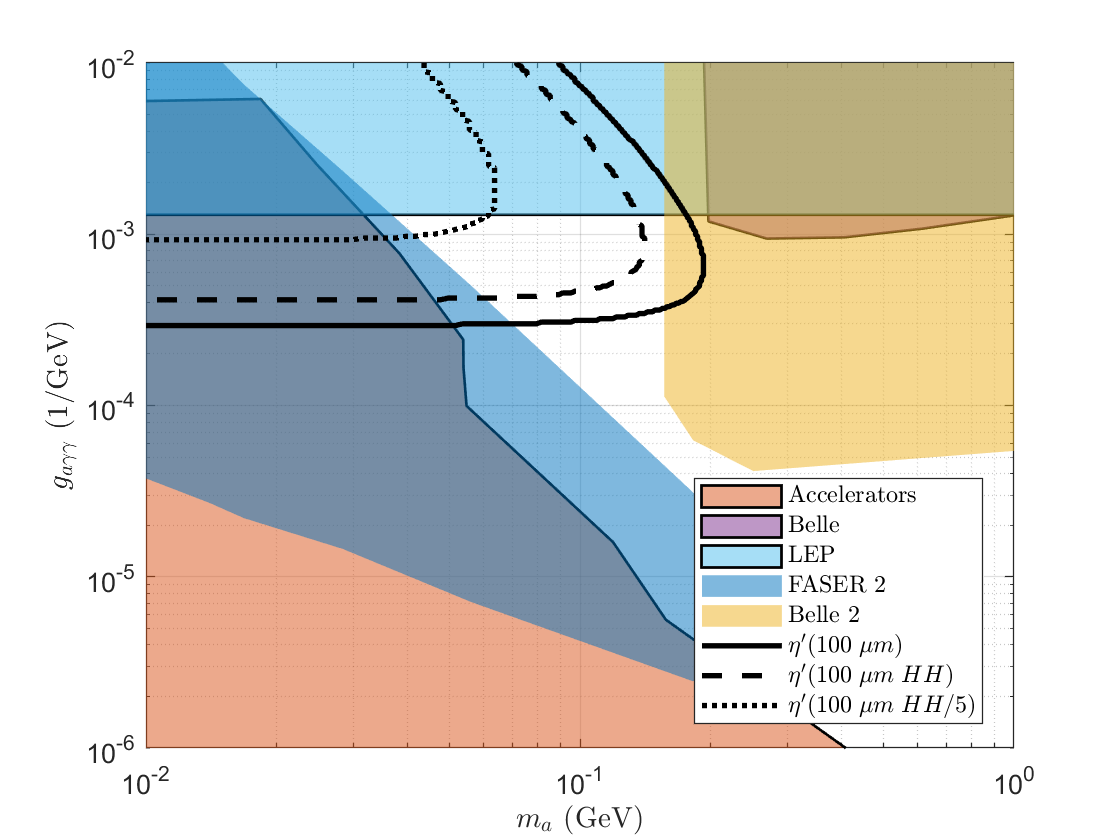} }
\caption{The regions to be probed at NICA after 1 year of operation at 95\% CL with ALPs produced in $\eta'$ meson decays. Left plot shows results for different minimal recognizable travel distance of ALP, parameterized by $L_{min}$ in eq.\,\eqref{Prob_2}. Right plot shows results for different values of $\Gamma_{tot}$. The existing limits (colored and
outlined) are taken from Belle at 95\% CL \cite{PhysRevLett.125.161806}, LEP at 95\% CL \cite{LEP}, accelerator experiments (NA64 at 90\% CL \cite{PhysRevLett.125.081801}, E137 at 95\% CL \cite{PhysRevD.38.3375}, NuCal at 90\% CL \cite{NuCal_alp}) and expected reaches of the ongoing experiments (colored) are given for FASER at 95\% CL \cite{PhysRevD.98.055021}, Belle-II at 90\% CL \cite{Dolan2017}.} 
\label{fig:Axion_from_etaprime}
\end{figure}

For ALPs we use the same simulations of ion collisions as we exploited above for the model with $A'$. The squared amplitude of process depicted in Fig.\ref{fig:Diagram_axion} is calculated using CalcHEP package \cite{Belyaev_2013} and then integrated over the interesting region of the phase space as explained in the case of $A'$. The achieved results are depicted in Fig.\ref{fig:Axion_from_eta} and Fig.\,\ref{fig:Axion_from_etaprime},  
where the black line refers to 3 signal events required within the Poisson statistics to exclude the corresponding outlined regions at 95\% CL (there is no dangerous background for the displaced vertex of two photons from the ALP decay). 
We find that the pion contribution to the ALP production is negligible for the model parameters in the previously unexplored regions. 

\begin{figure}[!htb]  
\centerline{\includegraphics[width=0.5\linewidth]{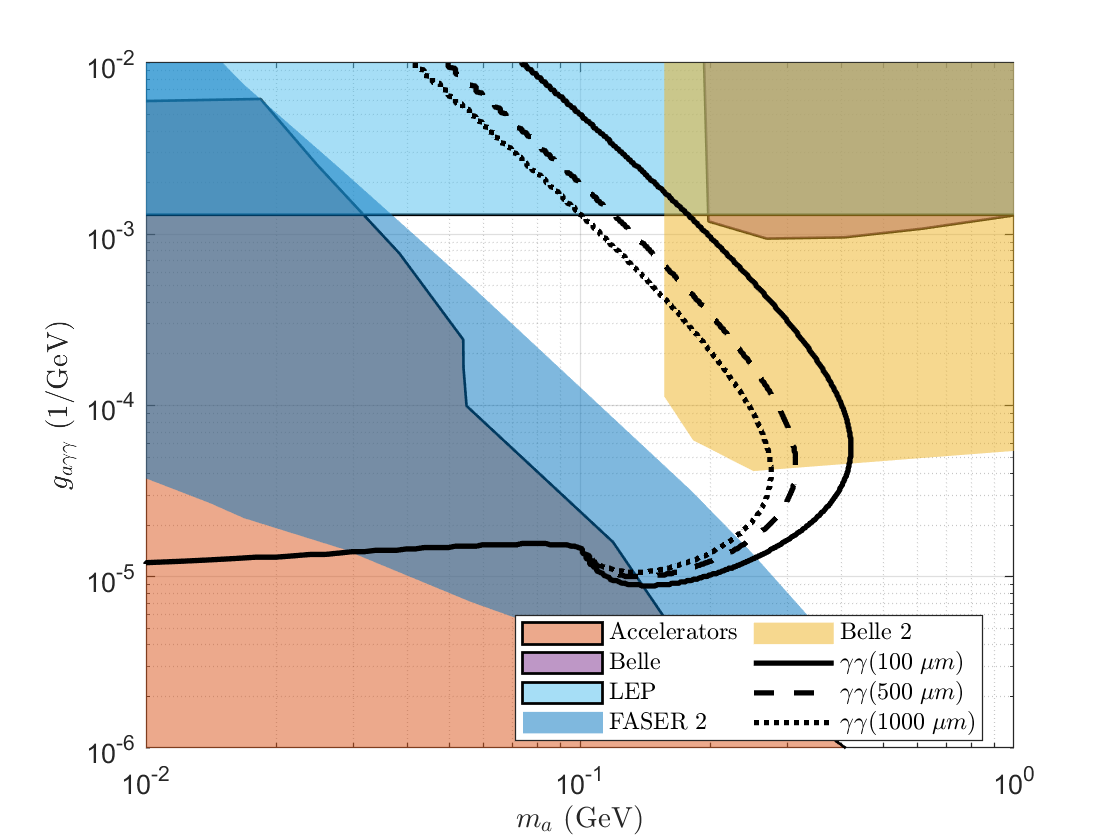} 
    \includegraphics[width=0.5\linewidth]{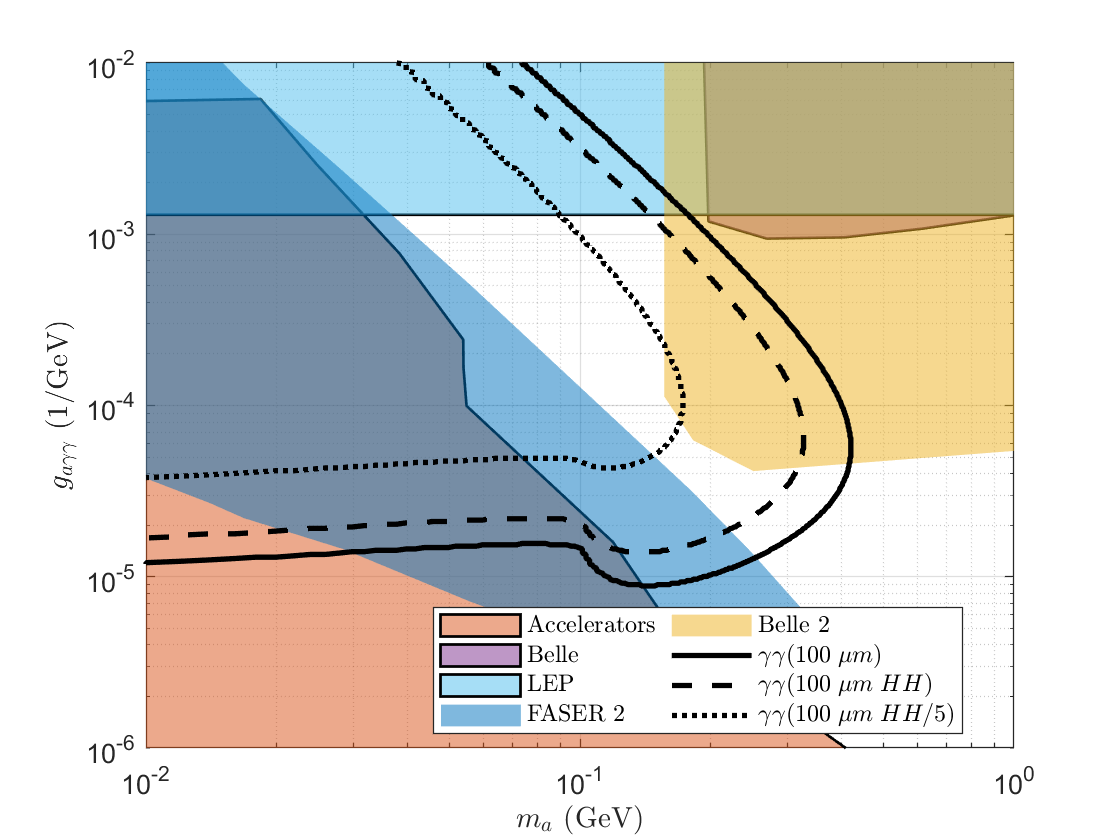} }
    \caption{The regions to be probed at NICA after 1 year of operation at 95\% CL with ALPs produced in ultraperipheral collisions. The existing limits (colored and
outlined) are taken from Belle at 95\% CL \cite{PhysRevLett.125.161806}, LEP at 95\% CL \cite{LEP}, accelerator experiments (NA64 at 90\% CL \cite{PhysRevLett.125.081801}, E137 at 95\% CL \cite{PhysRevD.38.3375}, NuCal at 90\% CL \cite{NuCal_alp}) and expected reaches of the ongoing experiments (colored) are given for FASER at 95\% CL \cite{PhysRevD.98.055021}, Belle-II at 90\% CL \cite{Dolan2017}.}
    \label{fig:Direct}
\end{figure}

There is also a direct contribution to the ALP production in heavy ion collisions due to rescatterings of the secondary photons abundantly produced in the collisions and subsequent relaxation processes.  
To estimate the number of ALPs coming from that photon-photon scattering we make use of the  method presented in \cite{Vysotsky:2019qou} and  perform calculations for Bi\,Bi collisions at the same luminosity of $L=5 \cdot 10^{27}\,\text{ cm}^{-2}\,\text{s}^{-1}$ and center-of-mass energy, $\sqrt{s_{\text{NN}}} = 9.2$\,GeV, as we adopted for simulation of the light meson production described above. In our calculations we use the monopole approximation for equivalent photon spectrum:
\begin{equation}
    n(\omega) = \frac{Z^2 \alpha}{\pi} \left[ \l 2\frac{\omega^2}{(\Lambda \gamma)^2} + 1 \r \ln \l 1+\frac{(\Lambda \gamma)^2}{\omega^2 } \r -2 \right] \frac{1}{\omega},
\end{equation}
where $\omega$ is the photon energy, $\gamma$ is the ion gamma-factor, and we set $\Lambda = 50$\,MeV. Then we obtain for the  cross section of $NN \rightarrow NNa$:
\begin{equation}
    \sigma = \sigma(\gamma \gamma \rightarrow a) \int_{m_a^2/\omega_{max}^2}^{\omega_{max}^2/m_a^2} \frac{dx}{8} n\l \sqrt{\frac{m_a^2x}{4}} \r n \l \sqrt{\frac{m_a^2}{4x}} \r,
\end{equation}
with photon energy ratio $x \equiv \omega_1/\omega_2$ and $\omega_{max} \gg \Lambda\gamma$ solving the equation $n(\omega_{max})=0$. 
Here again we require that energies of the photons from ALP decays exceed $50$\,MeV. The corresponding to 3 signal events contours are presented in Fig.\,\ref{fig:Direct}. 
One observes from Figs.\,\ref{fig:Axion_from_eta}-\ref{fig:Direct}, that the direct production exhibits similar to $\eta$-meson channel potential in testing the models with light ALPs. 

{\bf 5.} To summarise, in this letter, we start the investigation of the NICA perspectives in searches for hypothetical light particles. We study models with light vectors and model with light axion-like particles. After an upgrade of MPD with ITS it becomes possible to use the displaced vertex as a signature of the light particle decays. It allows to cover yet unexplored regions of new physics model parameter space and cross check results of other ongoing experiments. Implementing HH case in our analysis we further widen the spectrum of models which NICA can actually explore. One may anticipate good prospects for NICA in testing models with light scalars and axial vectors as well. 

Our results for ALPs may be improved by taking into account fermionic couplings and non radiative decays of mesons to ALP inherent in particular models. However we don't expect a significant improvement in the  NICA sensitivity. Then, while the direct production is a  promising source of ALPs but there are known uncertainties in calculation of the ALP production in such processes, e.g. how to determine the number of ultra-peripherial collisions and the spectrum of viable photons. Still it seems to be the most powerful source of ALPs at high collision energies with higher $\gamma$-factor for ions. In the case of $A'$ it may be worth to conduct more detailed research of ultraperipheral collisions. As soon as the details (type of colliding ions, their energies, luminosity, ITS resolution and other) of MPD update are finally fixed one will refine the present analysis.

\vskip 0.3cm
We thank S.\,Demidov for stimulating ideas and S.\,Godunov for interesting discussions. We especially appreciate  V.\,Riabov for valuable comments on the manuscript, for the information on the NICA collider and detector systems and for sharing files with kinematics of mesons simulated for NICA beam scatterings. 
The work on the expected signal from decaying light exotic particles is supported by the Russian Science Foundation RSF grant 21-12-00379. The work of D.\,K. on the expected background is supported by the Foundation for the Advancement of Theoretical Physics and Mathematics “BASIS”. 

%%%%%%%%%%%%%%%%%%%%%%%%%%%%%%%%%%%%%%%%%%%%%%%%%%%%%%%%%%%%%%%%%%%%%%%%%%%%%%

\vskip 1cm

\bibliographystyle{utphys}
\bibliography{refs}
\end{document}